**Investigation of phonon behavior in $Pr_2NiMnO_6$ by micro-Raman spectroscopy**

**K.D. Truong, M.P. Singh, S. Jandl, and P. Fournier**

Regroupement québécois sur les matériaux de pointe, Département de Physique, Université de Sherbrooke, Sherbrooke, Canada, J1K 2R1

The temperature dependence of phonon excitations and the presence of spin phonon coupling in polycrystalline $Pr_2NiMnO_6$ samples were studied using micro-Raman spectroscopy and magnetometry. Magnetic properties show a single ferromagnetic-to-paramagnetic transition at 228 K and a saturation magnetization close to 4.95 $\mu_B$/f.u.. Three distinct Raman modes at 657, 642, and 511 $cm^{-1}$ are observed. The phonon excitations show a clear hardening due to anharmonicity from 300 K down to 10 K. Further, temperature dependence of the 657 $cm^{-1}$ mode shows only a small softening. This reflects the presence of a relatively weak spin-phonon coupling in $Pr_2NiMnO_6$ contrary to other double perovskites previously studied.

## I. Introduction

Research interest in $R_2NiMnO_6$ (RNMO, R = rare earth cations) double perovskites arises owing to their various technologically important properties, such as ferromagnetic insulators, spin-phonon, and magnetoelectric coupling [1-8] These complex physical properties are the result of competing interactions between magnetic (spins), structural (phonons), and polarization (charges) order parameters. The possibility of tuning the magnetoelectric [3-5] and spin-phonon coupling [9, 10] through meticulous modifications in their structural parameters is a step forward for their potential uses in spintronic devices. Additionally, these intricate modifications offer an opportunity to investigate the underlying mechanisms that govern the coupled properties.

Recent works, in this class of compounds, have been primarily focused on $La_2NiMnO_6$ (LNMO). They include the synthesis conditions that promote the Ni/Mn cation ordering and its impact on the physical properties [6]. In particular, magnetic properties are found to be sensitive to the Ni/Mn ordering. Hence, an ordered LNMO phase shows a single magnetic transition at about 280 K, while a partially ordered LNMO phase shows an additional transition at about 138 K [6]. Thus, these properties have been used as a tool to determine the structural quality of a given sample [9-12]. Despite the extensive work on LNMO [2-12], very little is known about the physical properties of other members of the RNMO family. In this context, Booth *et al.* [13] have recently synthesised a series of RNMO samples in polycrystalline form and explored their structural and magnetic behaviours. This study shows that the size of the rare earth element influences significantly the structural and magnetic properties of RNMO subject to the intricate modifications of the Ni-O-Mn bond lengths and bond angle. These modifications are expected to influence significantly the electronic and magnetic ground states and subsequently modify the phonon dispersion, the magnetic interaction strength, and eventually the spin-phonon coupling.

Apart from $La_2NiMnO_6$ [9, 10] no comprehensive information about the phonon behavior and spin-phonon coupling of other members of RNMO are available in the literature.

This has prompted to the present investigation of the temperature dependence of the phonon excitations and the spin-phonon coupling of $Pr_2NiMnO_6$ (PNMO) using polarized micro-Raman spectroscopy. We compare its physical properties with LNMO and find a relatively weak spin-phonon coupling in PNMO. We discuss the possible impact of the Pr ionic radius on the observed physical properties.

**II. Experiments**

Bulk $Pr_2NiMnO_6$ (PNMO) samples were synthesized using $Pr_6O_{11}$, NiO and $MnO_2$ as starting materials. First, the $Pr_6O_{11}$ were precalcined in air at 900°C. Stoichiometric quantities of these oxides were then mixed, grinded and poured into an alumina crucible. The mix was subjected to several heating cycles between 1000 °C and 1300 °C in air with various intermediate grinding steps to obtain a homogeneous mixture. In the final step, the resultant PNMO powder was pressed into pellets and annealed at 1450°C in air overnight. Although the sample was polycrystalline, the size of the constituting microcrystals was larger than the micro-Raman laser spot (see below).

The temperature dependence of the magnetization (M-T) in 500 Oe applied magnetic field and its field dependence (M-H loops) at 10 K were measured using a superconducting quantum interference device (SQUID) magnetometer from Quantum Design. Phonon spectra with the 0.5 $cm^{-1}$ resolution were recorded using micro-Raman spectrometer in the backscattering configurations with the 632.8 nm He-Ne laser line and a liquid-nitrogen cooled CCD detector [9]. Laser spot size was kept at 3µm. Such diameter is perfectly adapted to study polycrystalline samples with phonon symmetry selection rules [9-10]. The size of the studied microcrystals (~ 10 µm) was relatively large enough to warrant that phonons are not affected by boundary conditions. Further, the grain boundary and strain effects manifest usually in samples few hundred angstroms large. Thus, such effects will be minimal in our experiments. Moreover, we like to point that depolarization effect which is likely the case studying this experimental configuration could become important due to the shape of the microcrystals as compared to large single crystals. For the purpose of our experiments, the notch filter was set to a down limit of about 200 cm$^{-1}$. Polarized spectra were measured in the HH (*i.e.,* the incident and scattering radiation are parallel) and HV (*i.e.,* the incident and scattering radiation are perpendicular) configurations. In order to avoid sample heating, a laser spot with 0.3mW exciting power was used.

**III. Results and Discussion**

Crystallinity and phase formations of our samples were studied using a Philips powder x-ray diffractometer operated in a θ-2θ mode and equipped with a Cu-Kα radiation. Typical powder x-ray diffraction (XRD) pattern of a PNMO sample is presented in Fig1a. It is characterized by a series of sharp peaks revealing that samples are polycrystalline in nature. These patterns are indexed based on the monoclinic P2$_1$/n symmetry [13]. The extracted lattice parameters are a = 5.54 Å, b = 5.50 Å, and c = 7.70 Å. These values are found to be close the reported values in the literature [13]. Thus, XRD studies demonstrate that these samples are polycrystalline and possess a monoclinic symmetry.

A typical M-H loop of PNMO sample, measured at 10K, is presented in Figure 1b. It shows that the PNMO sample is characterized by a well-defined hysteresis and a value of saturation magnetization as large as 4.95 μ$_B$/ f.u. at 50 kOe field. This demonstrates that PNMO is ferromagnetic in nature. Although these samples show a coercivity of 450 Oe, the magnetic moments get fully saturated only in a large value of applied magnetic field (5 to 10 kOe). This shows that parts of the magnetic domains are probably pinned due to the structural defects. In both ordered Ni$^{2+}$ (d$^8$: t$_{2g}^6$ e$_g^2$) and Mn$^{4+}$ (d$^3$: t$_{2g}^3$e$_g^0$) and disordered low-spin Ni$^{3+}$ (d$^7$: t$_{2g}^6$ e$_g^1$) and high-spin Mn$^{3+}$ (d$^4$: t$_{2g}^3$e$_g^1$) configurations, the theoretical values of the spin-only saturation magnetization is about 5 μ$_B$/f.u. [3, 6, 7]. This is very close to the observed value of the saturation magnetization in our PNMO samples. Thus relying on the measured value of saturation magnetization, it remains difficult to determine the possible oxidation states of Ni/Mn in PNMO as was done for La$_2$CoMnO$_6$ (LCMO) [6, 7]. To get further insights, we measured the M-T curve (inset of Fig 1b). The M-T data shows that magnetization is independent of temperature in the low temperature confirming that PNMO is ferromagnetic. It shows also that the sample is characterized by a single ferromagnetic-to-paramagnetic magnetic transition around 228

K. Furthermore, no secondary magnetic transitions were observed in our samples at low temperature (~ 140 K) as often seen in the disordered LNMO and LCMO systems [6, 11, 12]. Thus, the absence of any additional magnetic transition at low temperature suggests that our PNMO samples are structurally ordered.

By drawing a parallel between LNMO and PNMO, the value of FM-$T_c$ at 228 K in PNMO can be assigned to the $Ni^{2+}$-O-$Mn^{4+}$ superexchange interaction modulated by relatively weaker amplitude of the superexchange strength [3, 6, 7]. Within the 180° superexchange framework [7], ferromagnetic transition temperature is determined by the magnitude of the spin-transfer integral that is governed by the overlap of Ni/Mn-O wave functions. In other words, it is controlled by the Ni-O-Mn bond length and angle. Its amplitude decreases exponentially with an increase in the bond length. In polycrystalline samples [13], the average Ni-O and Mn-O bond lengths in LNMO are 2.02Å and 1.917Å respectively, and the average Ni-O-Mn bond angle is 160.4°. On the contrary, the average Ni-O and Mn-O bond lengths in PNMO are 2.029Å and 1.928Å respectively, and the average Ni-O-Mn bond angle is 158.6°. This relative enhancement in the total bond length of Ni-O-Mn and the decrease in the Ni-O-Mn average bond angle reduce the effective overlap of Ni/Mn-O electronic wave functions leading to a net decrease in the amplitude of the spin-transfer integral. Consequently, compared to LNMO, a lower magnetic transition temperature is expected in PNMO as observed.

Raman spectroscopy is a sensitive technique to study the spin-phonon coupling in strongly correlated oxides [14-20]. It has indeed been successfully used to study the phonon behaviors and spin-phonon coupling in a variety of perovskite oxide materials [14-20], including double perovskites (*e.g.,* LNMO and LCMO) [9, 10]. The Raman spectra at room temperature of PNMO micro-crystals in the HH and HV polarization configurations are shown in Fig. 2a. For comparison, the Raman spectra of LNMO, measured in the same configurations, are also presented in Fig.2b. It is important to note that both oxides exhibit similar Raman spectra. Two sharp phonon excitations in both polarization configurations are observed and the relative intensity of these phonon excitations depends on the polarization configurations. Despite these global similarities, the phonon features significantly differ from each other. In PNMO, these excitations are observed at 657 cm$^{-1}$ and 511 cm$^{-1}$ in the HH configuration, and at 642 cm$^{-1}$ and 511cm$^{-1}$ in the HV configuration. In LNMO, these excitations appear at 671 cm$^{-1}$ and 530 cm$^{-1}$ for the HH configuration and at 668 cm$^{-1}$ and 530 cm$^{-1}$ for the HV configuration. Since LNMO, LCMO and PNMO have identical crystal structures, we use the LNMO and LCMO theoretical group analysis of Iliev *et. al.* [10] to interpret the observed phonon excitations in PNMO samples. The observed phonon modes at 657 cm$^{-1}$ and 511 cm$^{-1}$ can be associated to the stretching and antistretching vibrations of the (Ni/Mn)O$_6$ octahedra respectively.

In order to study the impact of temperature on the phonon behavior of PNMO, we measured the Raman spectra of our samples down to 10 K. The phonon spectra are presented in Figure 3. For the sake of comparison with LNMO and LCMO, we focus our current study on strong intensity phonons of PNMO that are affected by the spin-phonon coupling. These data show that the intensity of the phonon excitations increases significantly while their line width decreases as we progressively cool down the samples. This reflects the reduced phonon scattering at low temperature and enhanced Raman tensors possibly due to reinforced orbital polarization. Also, the frequency of these phonons is showing clear temperature dependence. There are two important sources for the origin of this temperature dependence: *(i)* anharmonicity [14], and *(ii)* spin-phonon coupling [15]. Anharmonicity effects are reproduced by the following relation $\omega(T) = \omega_o - C[1+2/(e^{\hbar\omega_o/2kT}-1)]$ [14]. Here $\omega_o$ and C are adjustable parameters. Usually, as one cools down the sample, phonon frequencies harden and approach a plateau at low temperature. As one may note, both stretching and antistretching modes (Fig 4) display a hardening down to 100 K. This illustrates that anharmonic effects are one of the dominant contributions in the observed temperature dependence of PNMO's phonon modes.

Within the mean-field approximation, the onset of long-range magnetic order induces a renormalization of the phonon frequency $\delta\omega_{sp-ph}$ (T) [15] proportional to (M(T)/M$_O$)$^{2,}$ where M(T) is the magnetization at temperature T and M$_O$ is the magnetization at 0 K. The relative shift in the temperature dependence of the symmetric excitation mode at 657

cm$^{-1}$ in the HH configuration is shown in Fig. 3a. In contrast to LNMO and LCMO, the onset of its softening is roughly at 100 K, which is far below PNMO's ferromagnetic-to-paramagnetic transition point at 228 K. Moreover it is small ~ 1.5 cm$^{-1}$ in comparison to the corresponding modes in LNMO and LCMO which can be as large as 7 cm$^{-1}$ from the ferromagnetic transition down to the low temperatures (~ 10 K) [9, 10]. Our experimental study reveals that the strength of the spin-phonon coupling in PNMO is weaker than in LNMO and LCMO. This suggests that in PNMO the force constants are less influenced by the modulation of the exchange interaction between Ni/Mn-O ions [20]. Nevertheless, our study shows a significant impact of the size of the rare earth cation on the spin-phonon coupling and phonon behavior of RNMO likely via a change in the Ni-O-Mn bond length and angle [20]. Studying the temperature dependence of the structural properties is warranted to understand the origin of this relatively weak spin-phonon coupling in PNMO.

## IV. Summary

In summary, we have studied the temperature dependence of the phonon behaviours and the spin-phonon coupling of ferromagnetic PNMO. By comparing our results with related literature for LNMO, we found that it possesses a weak spin-phonon coupling and that the properties of RNMO double perovskites are strongly influenced by the type of rare earth cations inserted in the crystal structure.

**Acknowledgements**

We thank S. Pelletier and M. Castonguay for their technical assistance. This work was supported by the Canadian Institute for Advanced Research, Canada Foundation for Innovation, the Natural Sciences and Engineering Research Council (Canada), and le Fonds Québécois pour la Recherche sur la Nature et les Technologies (Québec),

## References

1. Prellier W, Singh M P, and Murugavel P 2005 *J. Phys.: Cond. Maters.* **17**, R803; Eerenstein W, Mathur N D, and Scott J F, 2006 *Nature* **442**, 759.

2. Singh M P, Truong K D, Fournier P, Rauwel P, Rauwel E, Carignan L, Menard D 2009 *J. Mag. Mag. Mater.* 321, 1743.

3. Rogado N S, Li J, Sleight A W, and Subramanian M A 2005 *Adv. Mater.* **17,** 2225.

4. Padhan P, Guo H Z, LeClair P, and Gupta A 2008 *Appl. Phys. Lett.* **92**, 022909.

5. Singh M P, Truong K D, and Fournier P 2007 *Appl. Phys. Lett.* **91,** 042504.

6. Singh M P, Truong K D, Jandl S, and Fournier P 2009 *Phys. Rev. B* **79**, 224421.

7. Goodenough J B 1963, *Magnetism and the Chemical Bond* (*Inter Science Publisher, New York, USA*) chapter 3; Dass R I, Yan J Q, and Goodenough J B, 2003 *Phys. Rev. B* **68**, 064415; Dass R I and Goodenough J B 2003 *Phys. Rev. B* **67**, 014401.

8. Bull C L, Gleeson D, and Knight K S 2003 *J. Phys. Cond. Maters.* **15**, 4927.

9. Truong K D, Laverdière J, Singh M P, Jandl S, and Fournier P 2007 *Phys. Rev. B* **76,** 132413. Truong K D, Singh M P, Jandl S, and Fournier P 2009 *Phys. Rev. B* **80,** 134424.

10. Iliev M N, Abrashev M V, Litvinchuck A V, Hadjiev V G, Guo H, and Gupta A 2007 *Phys. Rev. B.* **75,** 104118; Iliev M N, Guo H, and Gupta A 2007 *Appl. Phys. Lett.* **90**, 151914; Iliev M N *et.al.* 2009 *J. Appl. Phys.* **106,** 023515.

11. Singh M P, Grygiel C, Sheets W C, Boullay Ph, Hervieu M, Prellier W, Mercey B, Simon Ch, and Raveau B 2007 *Appl. Phys. Lett.* **91,** 012503.

12. Boullay Ph, Grygiel C, Rautama EL, Singh M P, and Kundu A K 2007 *Mat. Sc. Eng. B* **144**, 49.

13. Booth R J, Fillman R, Whitaker H, and Nag A, Tiwari AM, Ramanujachary K V, Gopalakrishnan J, and Lofland S E 2009 *Mater. Res. Bull.* **44**, 1559.

14. Balkanski M, Wallis R F, and Haro E 1983 *Phys. Rev. B* **28**, 1928.

15. Granado E, Garcia A, Sanjurjo J A, Rettori C, Torriani I, Prado F, Sanchez R D, Caneiro A, and Oseroff S B 1999 *Phys. Rev. B* **60**, 11879.

16. Zhao S, Shi L, Zhou S, Zhao J, Yang H, and Guo Y 2009 *J. Appl. Phys.* **106**, 123901

17. Dediu V, Ferdeghini C, Matacotta F C, Nozar P, and Ruani G 2000 *Phys. Rev. Letts* **84** 4489

18. Tatsi A, Papadopoulou E. L., Lampakis D, Liarokapis E, Prellier W, and Mercey B 2008 *Phys. Rev. B* **68**, 024432; M. V. Abrashev, J. Bäckström, and L. Börjesson M. Pissas N. Kolev and M. N. Iliev 2001, *ibid* **64,** 144429

**19.** Gupta R., Pai G V, Sood A.K., Ramakrishnan T.V. and Rao C. N. R. 2002 *Europhys. Lett.* **58** 778; Mansouri S, Charpentier S, Jandl S, Fournier P, Mukhin A A, Ivanov V Y and Balbashov A 2009 *J. Phys.: Condens. Matter* **21** 386004

**20.** Laverdière J, Jandl S, Mukhin A A, Ivanov V Y, Ivanov VG, and Iliev M N 2006, Phys. Rev. B **73**, 214301

**Figure captions**

**Figure 1: (Color online) (a)** Typical θ-2θ x-ray diffraction pattern of a PNMO sample indexed based on P2$_1$/n symmetry. **(b)** Magnetic field dependence of the magnetization (i.e. M-H loop) measured at 10K. Inset: magnetization as a function of temperature measured in a 500 Oe applied magnetic field.

**Figure 2:** Typical polarized Raman spectra of PNMO measured in (a) HH and (b) HV configurations. Raman spectra of LNMO measured under identical conditions are also included in **(a)** and **(b) for the comparison**. In both panels, the spectra are shifted vertically for clarity.

**Figure 3: (Color online)** Raman spectra of PNMO at various temperatures measured in (a) the HH and (b) the HV configurations. The dotted line in Fig 3a is a guide to the eye. Spectra are shifted vertically for clarity.

**Figure 4: (Color online)** Temperature dependence shift in the Raman modes at **(a)** 657 cm$^{-1}$ and **(b)** 511cm$^{-1}$. The solid red lines are fits to the anharmonic phonon effects (see text).

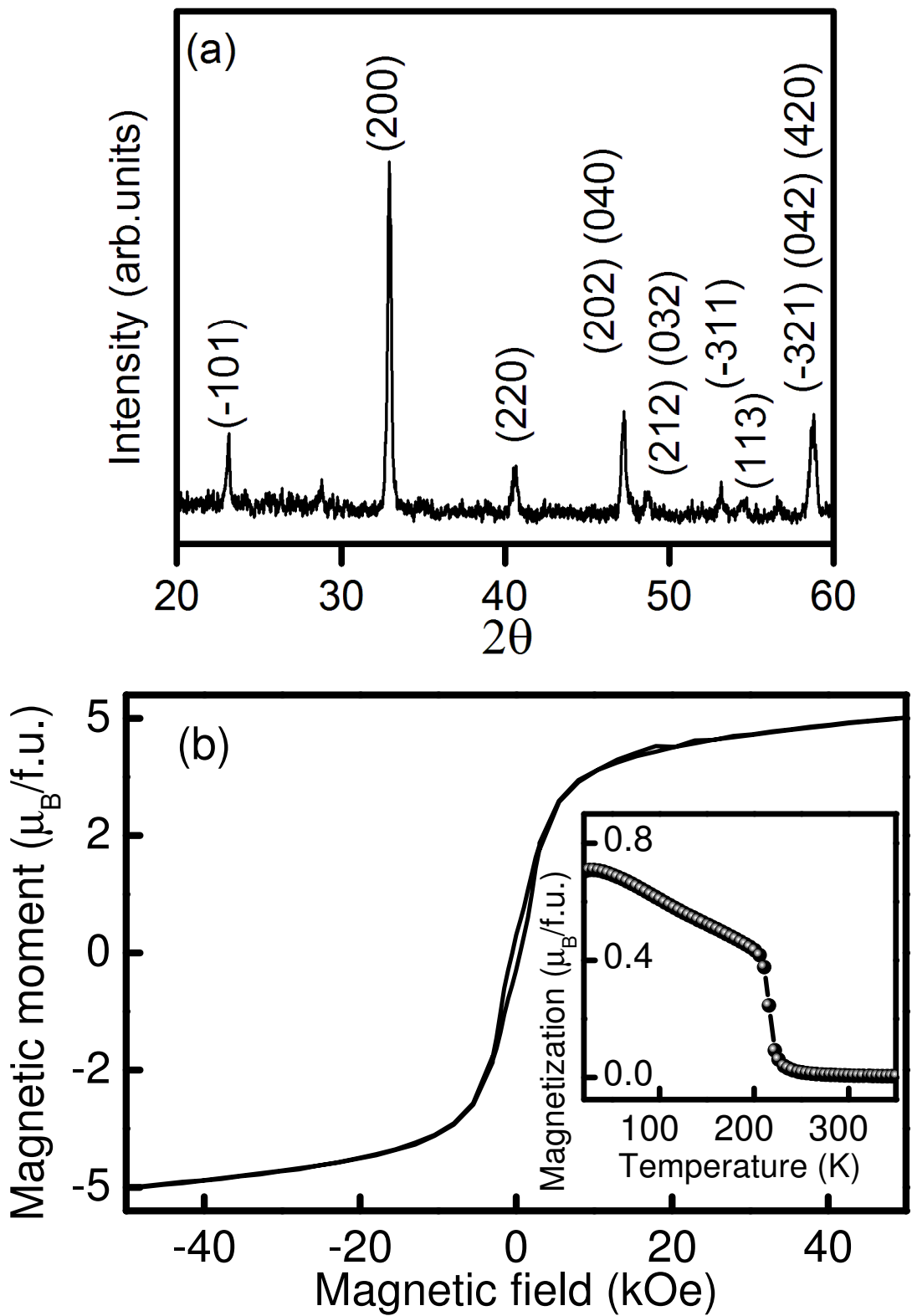

Fig. 1

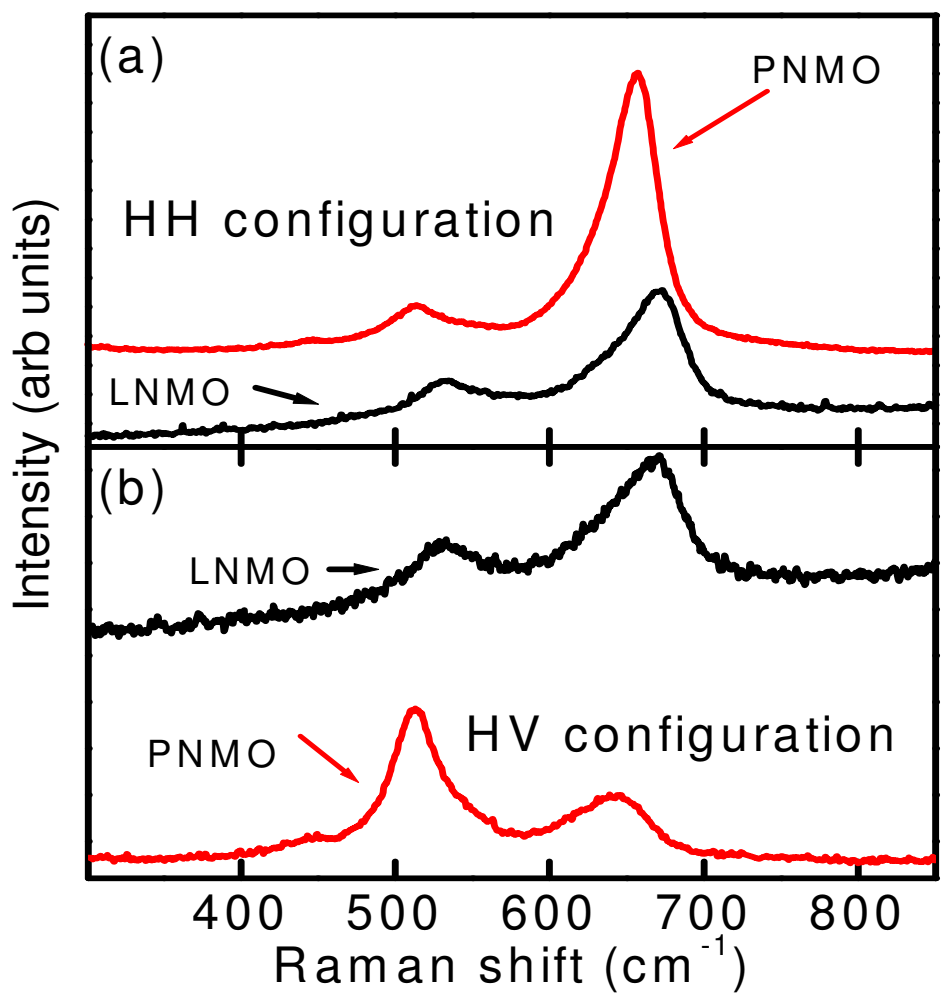

Fig. 2

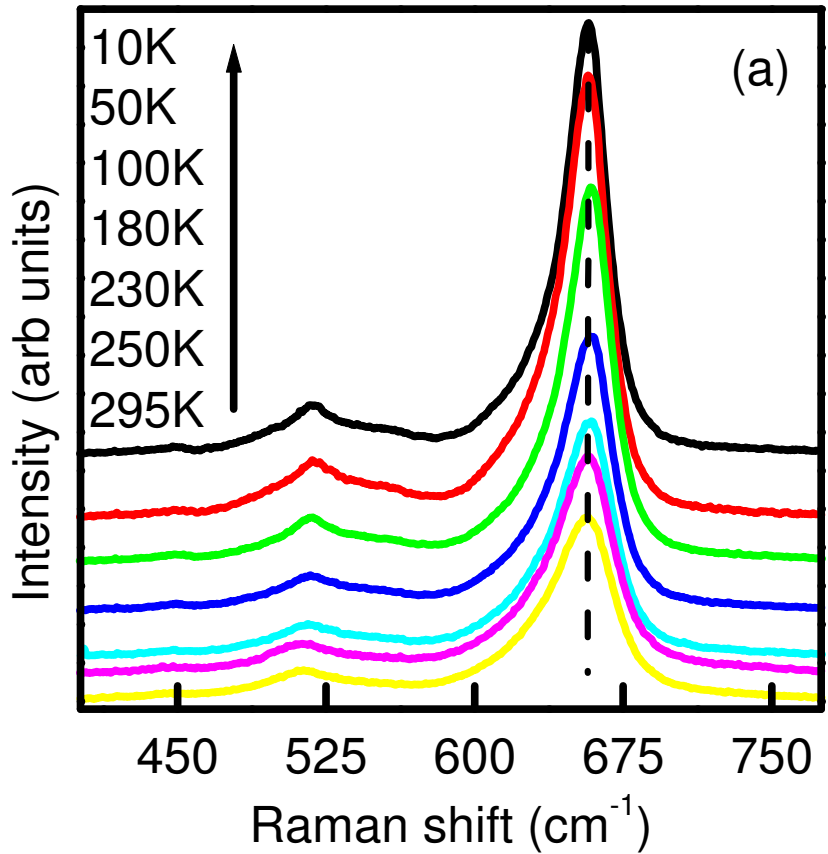
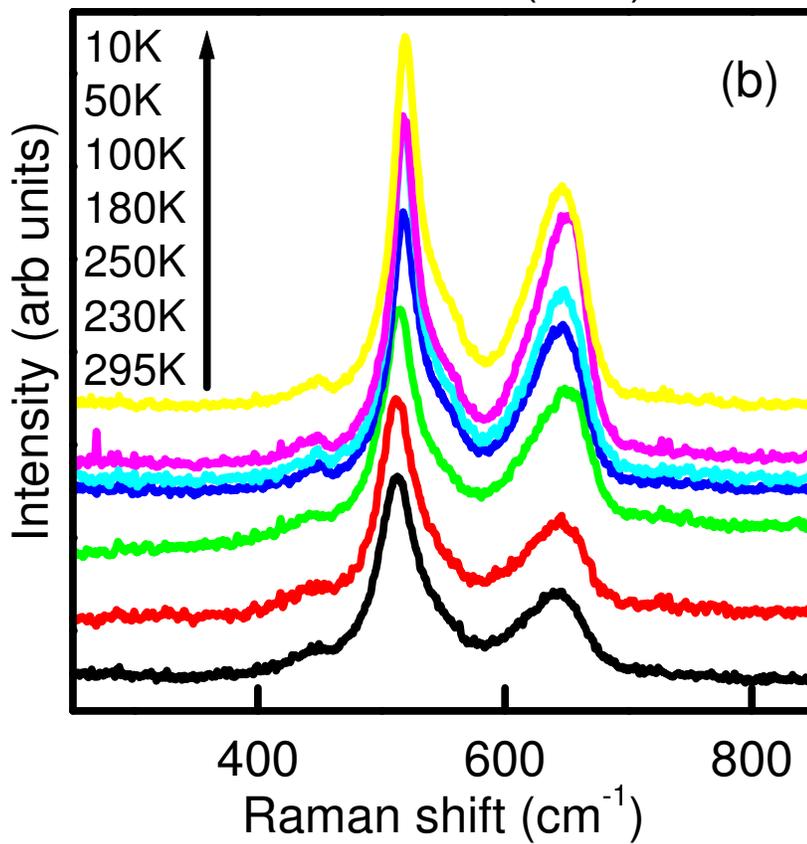

Fig. 3

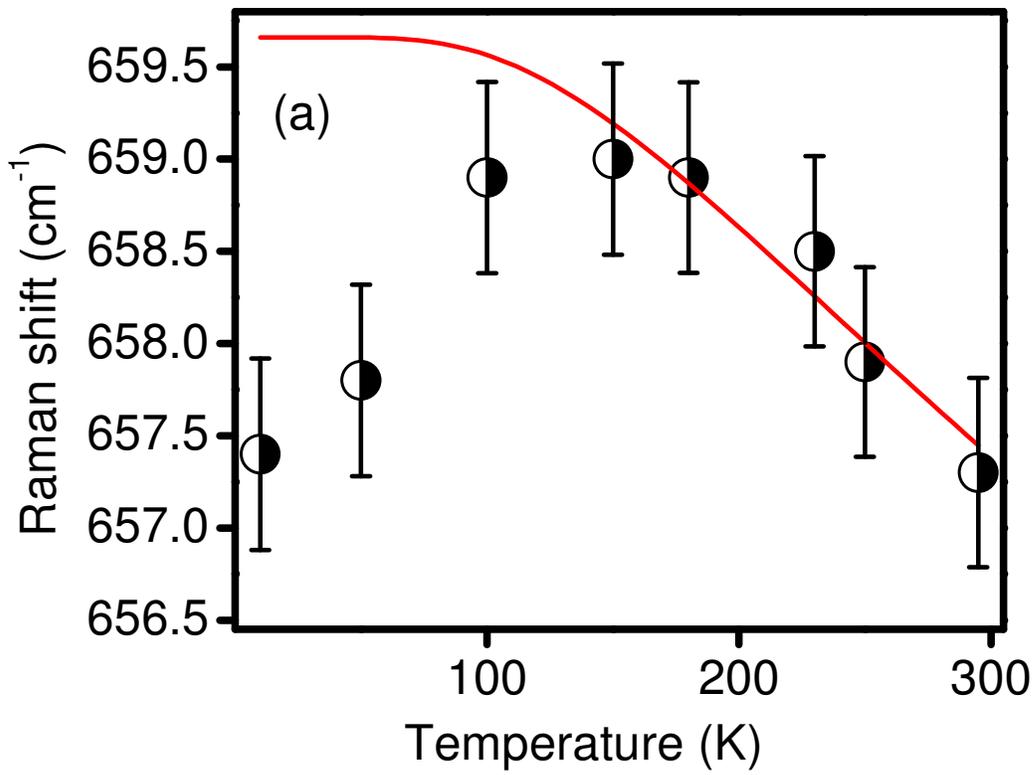

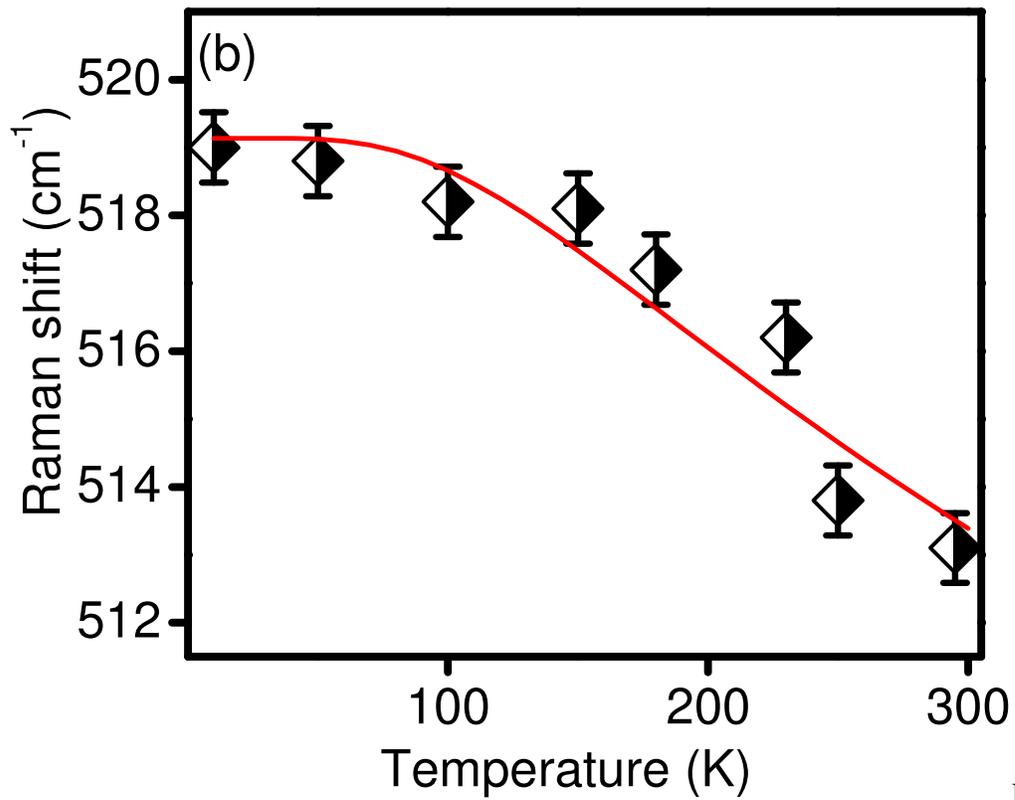

Fig. 4